\documentclass[multphys,vecphys]{svmult}

\usepackage{makeidx}         
\usepackage{graphicx}        
\usepackage{multicol}        
\usepackage[bottom]{footmisc}

\makeindex             

\begin{document}

\title*{First results from the VIMOS-IFU survey of gravitationally 
lensing clusters at $z \sim 0.2$}
\titlerunning{VIMOS-IFU survey of lensing clusters at $z\sim 0.2$: First results} 

\author{Giovanni Covone\inst{1,2}
\and Jean-Paul Kneib\inst{2}
\and Genevieve Soucail\inst{3}
\and Eric Jullo\inst{4}
\and Johan Richard \inst{5}}

\institute{INAF -- Osservatorio Astronomico di Capodimonte, Naples, Italy
\texttt{covone@na.astro.it}
\and OAMP -- Laboratoire d'Astrophysique de Marseille, France
\and OMP -- Laboratoire d'Astrophysique de Toulouse-Tarbes, France
\and European Southern Observatory, Santiago, Chile
\and California Institute of Technology, Pasadena, USA}
%
%
\maketitle

\begin{abstract}

We present the on-going observational program of a VIMOS Integral Field Unit 
survey of the central regions of massive, gravitational lensing 
galaxy clusters at redshift $z \simeq 0.2$.
We have observed six clusters using the low-resolution blue grism
($R \simeq 200$), and the 
spectroscopic survey is complemented by a wealth of photometric data,
including {\it Hubble Space Telescope} 
optical data  and near infrared VLT data.
The principal scientific aims of this project are: the
study of the high-$z$ lensed galaxies, the transformation and evolution of 
galaxies in cluster cores and the use of multiple images to
constrain cosmography.
We briefly report here on the first results from this project on the 
clusters Abell 2667 and Abell 68.

\end{abstract}

\section{Introduction}
\label{sec:1}

Because of their intense gravitational field, massive 
(i.e., $M > 10^{14} M_{\odot}$)
clusters of galaxies act as Gravitational Telescopes (GTs)
and are therefore an important tool to investigate the
high-redshift Universe (see, e.g., \cite{gt}).
In order to fully exploit the scientific potential of the GTs 
we have started an extensive integral field spectroscopy (IFS)
survey of massive galaxy clusters.
Targets have been selected among well-known gravitational lensing clusters 
between redshift $\sim 0.2$ and 0.3,
for which complementary {\it Hubble Space Telescope} (HST) 
data are available.
All the clusters are X-ray bright sources. 
Our sample partially overlaps with
the one analyzed in \cite{smith05}.

IFS is the ideal tool in order to obtain spatially complete spectroscopic
information of compact sky regions such as cluster cores.
Moreover, cluster cores are the regions where strong lensing 
phenomena are observed (i.e., giant arcs and multiply imaged sources): 
for clusters in our sample, strongly lensed galaxies are within
$\theta \simeq 1$ arcmin from the cluster center.

VIMOS-IFU \cite{lefevre03} is thus the natural choice for such observational 
program, since, at present, it provides the largest f.o.v. for among 
integral field spectrographs mounted on the 8-10m telescopes.

All the clusters in our sample have been observed using the 
low-resolution blue (LR-B) grism, with a spectral resolution $R \sim 200$.
Taking into account the lower efficiency at the end of the spectra
and the zero order contaminations, the final useful spectral range 
is limited between $\simeq 3900$ \AA \, and 6800 \AA.
This spectral range is suitable both for detecting high-redshift source 
(e.g., Ly$\alpha$ emitters in the redshift range $2.2 < z < 5.5$ or 
[OII] emitters out to $z\sim 0.8$)
and to sample the rest-frame 4000 \AA \, break for the
cluster galaxy population.
With a fiber size 0.66 arcsec, 
the IFU f.o.v. covers a contigous region of $54\times 54$ 
arcsec$^2$, sampled by 6400 fibers.

A subset of the clusters in the sample 
has also been observed using a higher resolution grism ($R\simeq3000$) 
covering the $\lambda= 6300-8600 \, $ \AA\ range.
These observations are useful to probe 
higher redshift Ly$\alpha$ emitters at $4.2<z<6.1$
or [OII] emitters at 0.7$<z<$1.3.

Observations have been completed (see Table 1 in \cite{soucail06}),
and data reduction is now in progress.
The data reduction process has been described in \cite{covone06}, and
\cite{zanichelli05} gives details about VIPGI, the VIMOS dedicated pipeline.
Furthermore, we have developed a Sextractor-based tool 
to help in object detection and 
spectra extraction from the fully reduced datacube \cite{jullo05}.

Altogether, our IFS survey covers a region 
of about 9 square arcmin in the central regions of six massive clusters.

Hereafter, we briefly report on the first results from this project:
the mass distribution in Abell 2667 in Sect. \ref{sec:2}, 
the properties of a magnified high-$z$ source in Abell 68 in Sect. \ref{sec:3} 
and an investigation of cluster galaxies in Sect. \ref{sec:4}.
We refer to the presentation by G. Soucail \cite{soucail06} for a detailed
discussion of the cosmography aspect of the project.

\section{Mass distribution model: A2667}
\label{sec:2}

\begin{figure}
\centering
\includegraphics[height=5cm]{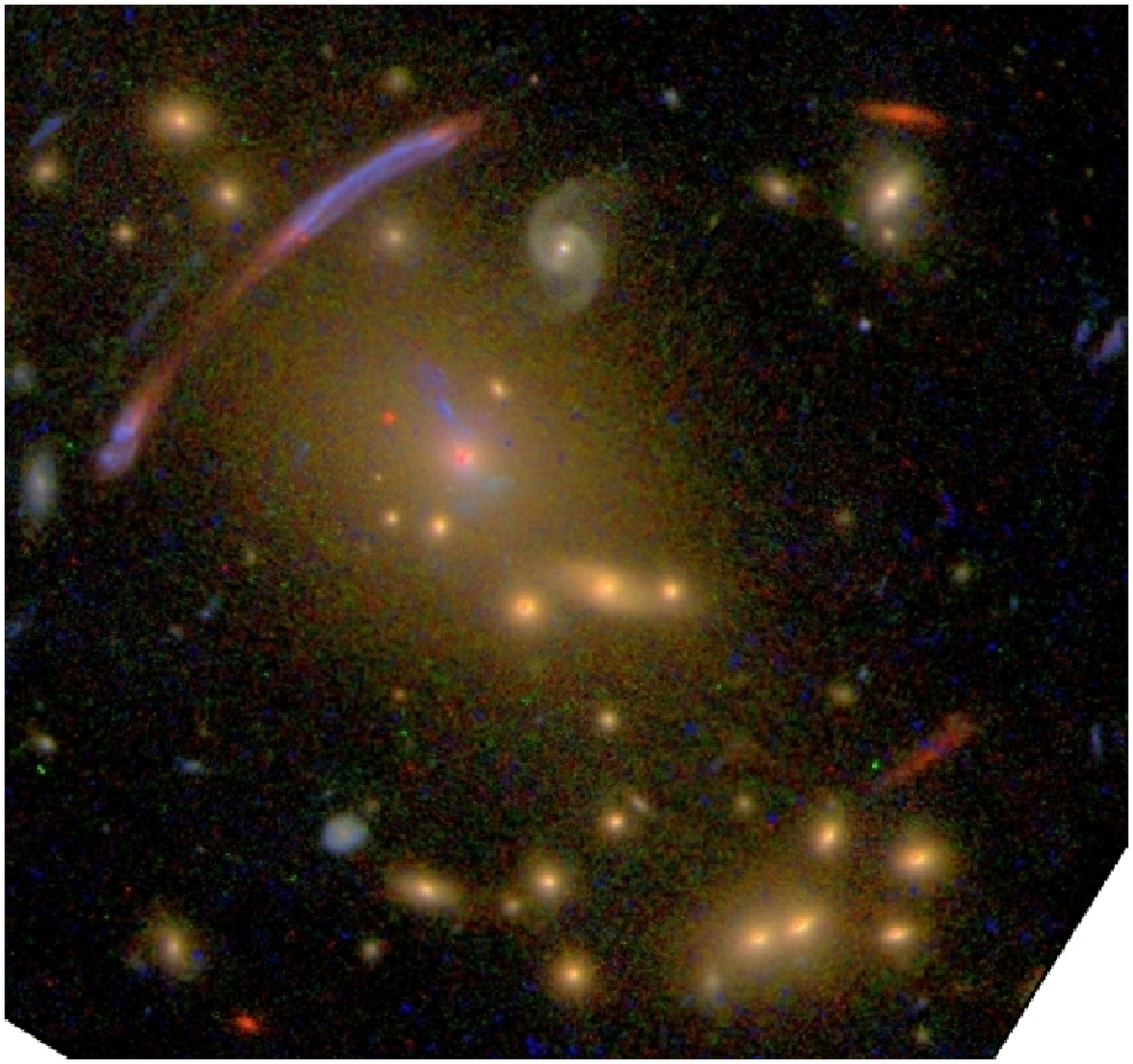}
\includegraphics[height=5cm]{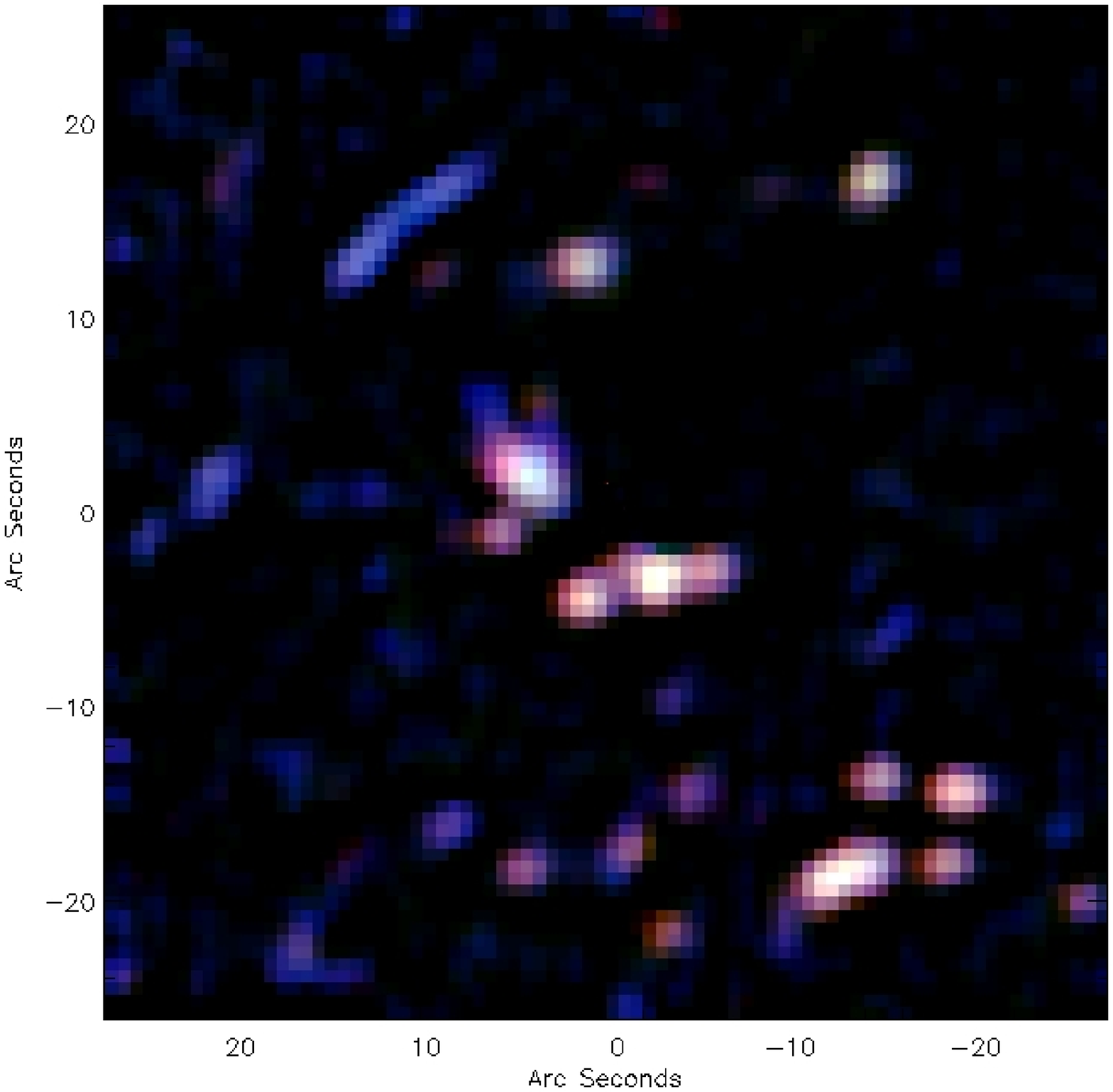}
%
\caption{Three color images of the inner arcmin square in the 
cluster A2667. Left panel: image created from the 
HST-WFPC2 data, filters 
B$_{450}$, V$_{606}$ and I$_{814}$.
Right panel: from the VIMOS-IFU datacube, using slices centered
at $\lambda=4600, 5200 $ and 6000 \AA.}
\label{fig:1}    
\end{figure}

Wide field IFS of the cluster central regions provides simultaneously
spectroscopic redshifts of both the cluster members and the images of the
gravitationally lensed sources, thus allowing a direct comparison of the strong
lensing analysis with the dynamical one.
Abell 2667 (hereafter, A2667) is a very remarkable galaxy
cluster at  $z=0.233$: it is among the top 5\% most luminous X-ray
clusters (at its redshift), 
and shows one of the brightest gravitational arc in the sky (Fig.~\ref{fig:1}).

A2667 has been observed with VIMOS-IFU during two separate nights 
(june 2003), with a total of 4 pointings of 2400s each, centered on the cD
galaxy, using the LR-B grism (see \cite{covone06} for details).
A small offset of about 2 arcsec was performed between the first and last two
pointings. Therefore, as consequence of the small number of pointings, a not
optimal dithering strategy and observations carried on two separate nights, the 
sky subtraction has not been optimal.

Nevertheless, we have obtained the spectroscopic measurements 
of the redshift for 34 sources in
the central $54\times 54$ arcsec$^2$ region of the cluster, corresponding to a
box of $200 \times 200 \, h_{70}^{-2} \, {\rm kpc}^2 $ at the cluster 
redshift\footnote{We use a cosmological model with $\Omega_{\Lambda}=0.7$ and $\Omega_m=0.3$}.
It includes in particulars:
 22 cluster members (i.e., all the cluster members brighter than
V$_{606}=23.2$, AB system) and the three separate
images of the giant gravitational arc ($z=1.0334$).

Using the spectroscopic redshift and the multiple images identified on the
HST-WFPC2
image, we have built a strong lensing model and performed 
a dynamical analysis of the cluster core, both resulting in mass 
of $\simeq 7.2 \, \times 10^{13} \, M_{\odot}$ within the central 
$110 \, h_{70}^{-1}$ kpc, and a velocity dispersion of $\simeq 950 {\rm km
s}^{-1}$, close to the value derived  
from the X-ray temperature (assuming that the cluster follows
the $\sigma-T$ relation). 

Such agreement supports the idea that A2667 cluster core is in a relaxed
dynamical state, as expected from its regular X-ray morphology (Rizza
et al. 1998).
Therefore, A2667 core appears to be dynamically
evolved, in contrast with the large fraction ($70 \pm 20 \, \%$) of
unrelaxed clusters with similar X-ray luminosity at similar redshift 
\cite{smith05}.

\section{Physical properties of a unusual lensed high$-z$ source}
\label{sec:3}

The combination of IFS and the large magnification provided from strong lensing
gives a unique opportunity to study in detail (i.e., the spatially resolved) the 
spectral properties of intrinsically low-luminosity source 
in the high$-z$ Universe,
(see, for instance, \cite{swinbank}).

\begin{figure}
\centering
\includegraphics[height=4.0cm]{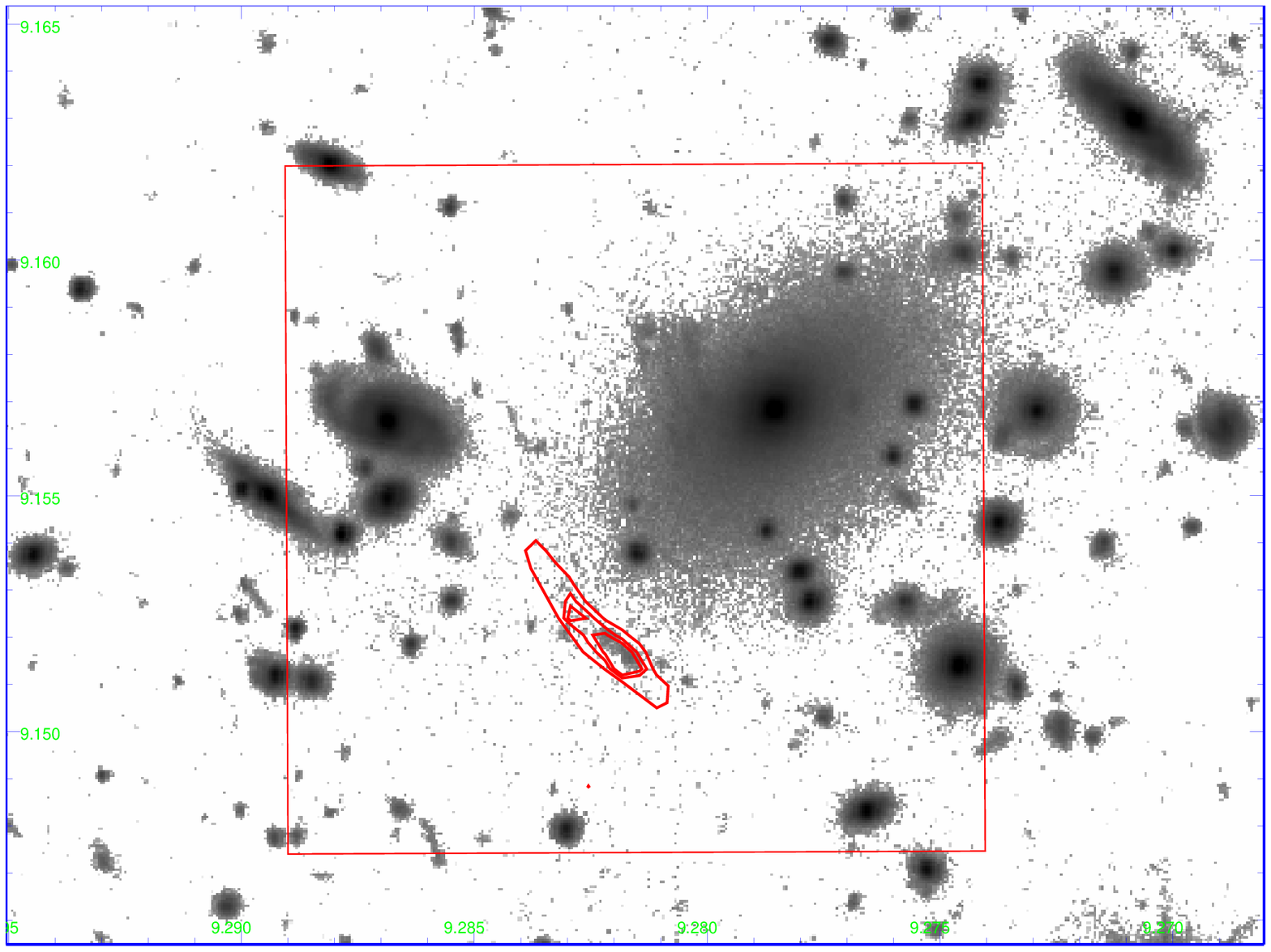}
\includegraphics[height=4.5cm]{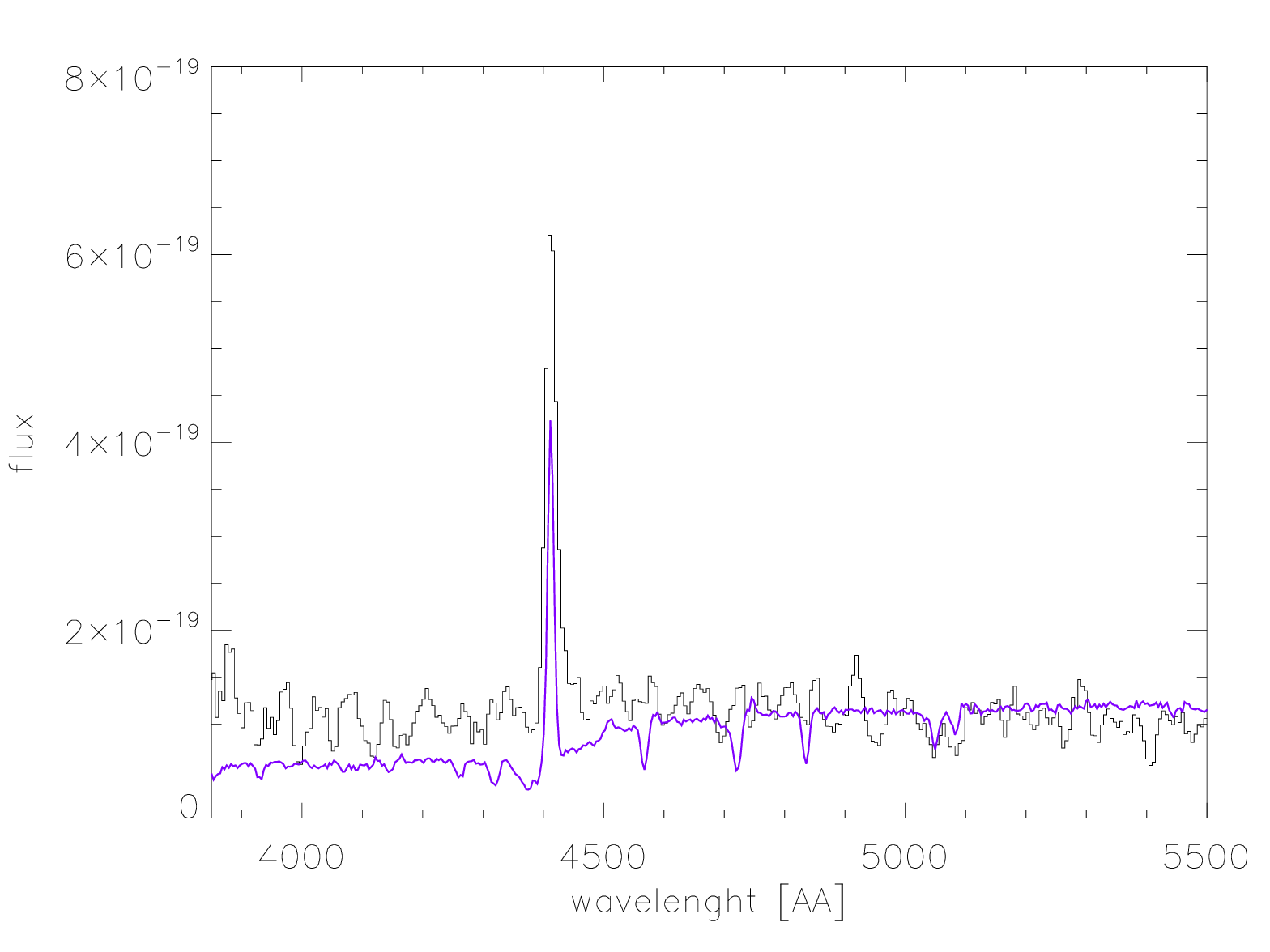}
\caption{Galaxy cluster A68. Left panel: HST R$_{707}$-band image of the
cluster, with the VIMOS-IFU f.o.v. and the lensed source C4. Note its wider
extension as detected by means of the IFS.
Right panel: one-dimensional VIMOS spectrum of the arc compared 
with a template of a $z=3$ Lyman-break galaxy from \cite{shapley}.}
\label{fig:2}    
\end{figure}

VIMOS-IFU observation of the core of the galaxy cluster Abell~68 
($z=0.255$) 
has revealed a surprisingly extended Ly$\alpha$ emission around a previously
known gravitationally lensed source \cite{covone06b} 
(denoted as C4 in \cite{smith05}, $z=2.625$). 
As shown in Fig.~\ref{fig:2}, the arc is seen to be $\sim 4$ arcsec in length 
on the HST-WFPC2 image (filter R$_{707}$, exposure time 7.5 ks). 
But in our shallow IFU pointing (4.8  ks), the arc is seen to be much
more extended \cite{covone06b}, reaching a maximal elongation of
$\simeq 10.8$ arcsec.
The emission line is not resolved in the IFU data, and its 
equivalent width is about $140 $ \AA , 
and remains constant within the errors along the source length.

%
According to the strong lensing model,
the source is single imaged and its magnification is 
$\mu\sim 35$: the intrinsic shape of the Lyman-$\alpha$ emitting region
has a disk-like appearance (see \cite{covone06b}) with a maximum 
extension of $\simeq 10 \, {\rm kpc}$. It is therefore about 
10 times smaller than Ly$\alpha$ blobs (see, for instance, 
\cite{matsuda04}) but much bigger
than a typical galaxy at $z\sim 2.5$
One possibility is that we are observing
a $\sim L_*$ galaxy undergoing a strong star-formation event,
with negligible dust-obscuration.

\section{Cluster galaxies investigations}
\label{sec:4}

\begin{figure}
\centering
\includegraphics[height=7cm, angle=270]{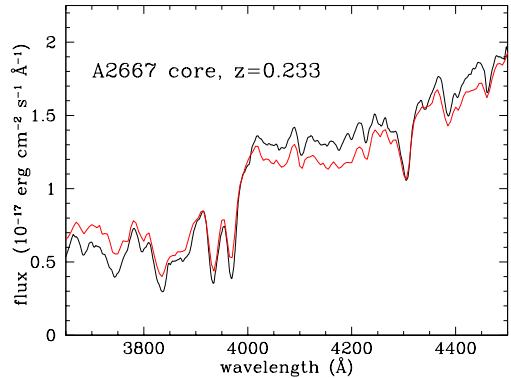}
\caption{Composite spectrum of the galaxy population in the core of the cluster
A2667 (black line), compared with the template spectrum of $z=0$ early-type galaxies
(red) from \cite{kinney}. The contribution of the cD galaxy is excluded, since
it appears to host an AGN.}
\label{fig:3}    
\end{figure}

The central region of rich galaxy clusters at intermediate redshift 
is expected to host an old and red
galaxy population, in great majority composed of early-type galaxies, see e.g.
\cite{kodama}, which at a redshift of $z<1$ is passively evolving, with
very low levels of star-formation activity. 

Recently, \cite{dressler} have shown that {\em composite cluster spectra}, 
built
from the light-weighted combination
of all the cluster members long-slit spectra, are a
useful tool to provide insights in the properties of the cluster population.
In this respect, IFS offer a unique possibility to build cluster 
composite spectra in unbiased way, since, for a given field of view, 
all clusters members are observed (without
any {\em apriori} selection) and, for each galaxy, all the light is collected
(therefore avoiding the possibility that for larger galaxies
some flux contribution might be
 missed due to the specific orientation of the slit).
Preliminary work (see Fig.~\ref{fig:3}) 
appears to confirm that the galaxy population in the 
very central region of A2667 is dominated by an evolved and passively aging
stellar population. 
We plan to build composite spectra for all the clusters, and to exploit the
recent {\em Spitzer} mid-infrared observations to complement this result and
provide quantitative upper limits on the hidden star-formation in the cluster
cores.

\end{document}